# Persistent Geospatial Outage Scenario Construction for Interdependent Infrastructure Simulation

Solongo Ganbold
Department of Computer Science
University of Nevada, Las Vegas
Las Vegas, Nevada, USA
ganbos1@unlv.nevada.edu

Gelila Webster
Department of Computer Science
University of Nevada, Las Vegas
Las Vegas, Nevada, USA
gelila.webster@unlv.edu

Sohini Roy
Department of Computer Science
University of Nevada, Las Vegas
Las Vegas, Nevada, USA
sohini.roy@unlv.edu

## ABSTRACT

Extreme weather events are producing persistent geographic patterns of power-grid disruption across the United States, yet these patterns are rarely incorporated into geospatial simulation of interdependent infrastructure failures. This paper presents a data-driven geospatial simulation framework that transforms persistent outage patterns into empirically grounded regional simulation scenarios for interdependent power-communication cascade simulation. Using a national outage dataset from 2015–2023, we introduce the Hotspot Persistence Index (HPI), a severity-ranked recurrence metric for identifying counties that repeatedly emerge as outage hotspots over time. We then apply a multi-scale density-based refinement process to convert persistent county-level hotspots into geographically interpretable regional simulation scenarios characterized by persistence, severity, and spatial extent. These empirically derived scenarios are used to parameterize the Modified Implicative Interdependency Model (MIIM), enabling simulation of cascading behavior across coupled power and communication layers. Results show that three persistent regional scenarios account for 54.4% of total HPI-weighted cascade impact, while communication-layer entities fail at 2.5× the rate of power buses under high-impact persistent scenarios. HPI-guided hardening reprioritizes protection candidates relative to degree- and betweenness-centrality baselines, demonstrating that empirically derived outage scenarios lead to different infrastructure protection priorities than topology-only approaches. These results demonstrate how empirically derived geospatial outage scenarios can improve simulation-based infrastructure resilience analysis and hardening prioritization.

## 1 Introduction

Extreme weather events are increasingly disrupting electric power infrastructure across the United States. Hurricanes, winter storms, heatwaves, wildfires, and extreme precipitation events can damage grid assets, interrupt distribution pathways, increase demand, and delay restoration. These disruptions are spatially uneven: some counties repeatedly experience severe outage activity, while others are affected only during isolated events. This makes persistence an important dimension of infrastructure vulnerability and an important input for simulation-based resilience analysis.

Prior work has examined outage frequency, weather-related impacts, community vulnerability, and spatial hotspot patterns, showing that outage risk is both geographically uneven and temporally dynamic. Separately, interdependent infrastructure models study how failures propagate across coupled systems such as power and communication networks. However, these two perspectives are rarely connected. Hotspot detection often stops at spatial characterization, while cascade simulations often rely on random, topology-based, or predefined initiating failures. As a result, persistent geospatial outage patterns are rarely transformed into empirically grounded failure scenarios for interdependent infrastructure simulation. This gap motivates our central question: can persistent outage regions be used to construct simulation-ready regional disruption scenarios, and do these scenarios lead to different cascade impacts and hardening priorities than topology-based approaches suggest?

To address this gap, this paper presents a data-driven geospatial simulation framework for interdependent infrastructure resilience analysis. Using a national outage dataset from 2015–2023, we first identify counties that repeatedly emerge as severe outage hotspots over time using the Hotspot Persistence Index (HPI). We then refine these county-level hotspots into geographically interpretable regional failure scenarios using a multi-scale density-based spatial clustering process. Finally, these empirically derived scenarios are used to parameterize the Modified Implicative Interdependency Model (MIIM) [1], enabling cascade simulation and hardening analysis in an interdependent power-communication network.

The key contribution of this paper is the transformation of historical geospatial outage observations into simulation-ready regional failure scenarios. The spatial analytics layer provides persistence, severity, and spatial-scope attributes, while the MIIM layer evaluates how those empirically grounded scenarios propagate through coupled power and communication infrastructures. In this sense, the paper advances geospatial simulation for infrastructure resilience by linking persistent outage patterns, regional scenario construction, and interdependent cascade evaluation, rather than treating outage hotspots as static map products.

This paper makes the following contributions:

1. **Persistence-aware outage scenario identification:** a Hotspot Persistence Index (HPI) for quantifying recurring county-level outage vulnerability over multiple years, supported by national and state-level robustness comparison.
2. **Geospatial outage scenario construction:** a multi-scale density-based spatial clustering process that converts persistent outage counties into geographically interpretable regional failure scenarios.
3. **Scenario-to-simulation parameterization:** a mapping procedure that translates regional scenario attributes, including persistence, severity, and spatial extent, into inputs for interdependent cascade simulation.
4. **MIIM-based cascade simulation and hardening analysis:** a simulation-based resilience evaluation of empirically derived outage scenarios, including cascading failure impact, communication-layer vulnerability, and infrastructure hardening priorities.

Together, these contributions provide a framework for moving from outage hotspot mapping to empirically grounded geospatial simulation for infrastructure resilience. Our results show that three persistent regional scenarios account for 54.4% of total HPI-weighted cascade impact, while communication-layer entities fail at approximately 2.5 times the rate of power buses under high-persistence scenarios. HPI-guided hardening also identifies protection priorities that differ from topology-only approaches, demonstrating that persistent geospatial outage patterns can support simulation-based infrastructure resilience analysis and more targeted hardening decisions.

**Figure 1: Overview of the proposed two-layer framework**

# 2 Related Work

Prior work relevant to this study falls into three areas: spatio-temporal hotspot detection and spatial clustering, climate-driven outage analytics, and simulation of interdependent infrastructure failures. These areas provide useful foundations, but they are often studied separately. Our work connects them by using persistent geospatial outage patterns to construct simulation-ready cascade scenarios for interdependent power-communication systems.

## 2.1 Spatio-Temporal Hotspot Clustering

Spatio-temporal hotspot detection is widely used to identify geographically concentrated event patterns and support planning, monitoring, and risk mitigation [2]. Density-based clustering is particularly useful in this context because it can identify irregularly shaped regions and treat isolated observations as noise. DBSCAN, introduced by Ester et al. [3], provides a useful baseline for identifying irregularly shaped spatial clusters and noise points in infrastructure disruption data.

However, national-scale outage data are spatially heterogeneous, and a single density threshold may either fragment meaningful regions or merge distinct patterns into overly broad clusters. Prior work on multi-density clustering highlights this challenge in urban and spatial event data [4, 5]. Temporal information adds another layer of complexity: recent hotspot studies have incorporated spatio-temporal and network constraints [6], extended DBSCAN-style approaches to spatio-temporal event data [7], and examined hotspot evolution and persistence using space-time methods [8]. We adopt a multi-scale density-based clustering process because it provides explicit control over the spatial scale of the resulting scenarios, which is necessary for constructing regional disruption units with interpretable geographic extent. These studies motivate our use of a persistence-aware, multi-scale clustering process for constructing outage simulation scenarios.

## 2.2 Climate-Driven Outage Analytics

Power outages are increasingly studied as indicators of infrastructure vulnerability under climate and extreme-weather stress. Severe weather can disrupt power systems through physical asset damage, flooding, generation disruption, distribution failures, and demand surges. Recent county-level outage studies show that weather-related outages are spatially uneven, temporally dynamic, and often concentrated in regions such as the Gulf Coast, Northeast, Great Lakes, West Coast, and Southern California [9–11].

Other work connects outage impacts with hazard exposure, socioeconomic vulnerability, and climate drivers. Poudyal et al. examine hurricane- and storm-surge-induced power-system vulnerability and socioeconomic impact [12], while event-specific and climate-driver studies analyze Hurricane Beryl-related outage disparities and La Niña-associated outage patterns [13, 14]. These studies demonstrate the importance of spatial outage analytics, but most focus on descriptive or predictive vulnerability characterization rather than transforming persistent outage patterns into inputs for downstream cascade simulation.

## 2.3 Interdependent Infrastructure Failures

Critical infrastructure systems are increasingly coupled through physical, cyber, and operational dependencies. Rinaldi et al. describe infrastructure interdependencies as a defining feature of modern systems, where failures in one sector can affect others and expand the consequences of disruptions [15]. Interdependent network studies further show that coupled systems may experience amplified cascading failures compared with isolated networks [16, 17].

Power and communication systems are especially interdependent because electric grids rely on communication networks for monitoring and control, while communication entities depend on power for continued operation. Prior work has modeled cascading failures in electric-cyber systems using Markov-chain approaches [18], mutually dependent power-communication networks [19], and topology-aware infrastructure risk models [20]. However, purely topological assumptions can be misleading because power-grid and communication failures propagate through operational dependencies as well as network structure.

In this work, we use the Modified Implicative Interdependency Model (MIIM) [1] as a downstream simulation layer. MIIM is suitable for this study because it represents power, communication, and cross-layer entities while capturing both reduced-operation and failed operational states.

## 2.4 Research Gap

Prior work provides strong foundations in hotspot detection, outage vulnerability analytics, and interdependent infrastructure modeling, but these areas are rarely connected. Existing hotspot studies identify spatial or spatio-temporal event concentrations, but they often do not focus on multi-year persistence of outage vulnerability. Existing outage analytics quantify frequency, duration, intensity, or regional vulnerability, but generally stop at descriptive or predictive characterization. Similarly, cascading-failure simulations evaluate system-level propagation, but their initiating disruptions are often random, topology-based, or predefined rather than derived from empirical outage persistence.

This paper bridges these gaps by transforming persistent outage hotspots into simulation-ready regional disruption scenarios. We introduce the Hotspot Persistence Index (HPI) to quantify recurring county-level outage vulnerability, apply multi-scale density-based spatial clustering to construct geographically interpretable regional scenarios, and use those empirically derived scenarios to parameterize MIIM-based cascade simulation. Thus, persistent outage hotspots are not only mapped, but converted into inputs for simulation-based infrastructure resilience analysis and hardening prioritization.

# 3 Dataset and Preprocessing

This study uses the OEDI Power Outage Dataset v2 (EAGLE-I), published by Pacific Northwest National Laboratory, covering county-level U.S. power outage records from 2015 to 2023. The cleaned analysis dataset contains 1,385,389 county-level outage records across 2,984 counties and 54 U.S. states and territories. Each outage record is associated with a geographic location, temporal attributes, and outage impact measures. The primary spatial unit of analysis is the county, which allows outage patterns to be compared consistently across states and regions while preserving sufficient geographic detail for national-scale geospatial scenario construction.

Although the raw dataset includes records from 2014, we exclude 2014 from the main analysis because coverage is incomplete, with records available only for November–December. Including 2014 in persistence calculations would artificially lower hotspot recurrence scores for counties with missing observations in earlier months. Therefore, the primary analysis uses the nine-year period from 2015 to 2023. The 2014 records are used only for exploratory context and are not included in the Hotspot Persistence Index (HPI), multi-scale density-based clustering, or MIIM-based cascade simulation.

Each record includes county and state identifiers, FIPS code, geographic coordinates, outage duration, maximum customers affected, and temporal information such as start time, month, year, and season. County centroids are used as the geographic representation for spatial clustering. Temporal variables are derived from the outage start time to support yearly, monthly, and seasonal analyses. Seasons are defined as winter, spring, summer, and fall based on the calendar month of each outage record.

To capture both the duration and customer impact of an outage, we define a severity score for each outage record as:

$$Severity = Duration \times Max\ Customers\ Affected \qquad (1)$$

This severity score is used throughout the framework as the primary outage-impact measure. It allows shorter outage records affecting many customers and longer outage records affecting fewer customers to be compared using a common metric. For each county and year, outage records are aggregated to compute annual severity values, which are then used to identify yearly hotspot counties, calculate HPI, and support simulation scenario construction.

Before analysis, the dataset is cleaned to remove invalid or incomplete records. Records with missing county or state identifiers, missing geographic coordinates, missing duration, or missing customer-impact values are excluded. Duration and maximum-customer fields are converted to numeric format, and temporal fields are standardized to ensure consistent month, year, and season assignment. After preprocessing, the cleaned dataset supports temporal outage characterization and the three main analytical stages described in Sections 4–6: county-level hotspot persistence computation, regional scenario construction, and interdependent cascade simulation. Table 1 summarizes the key variables used in the study.

Figure 2 provides an overview of the temporal structure of the dataset. The yearly outage count shows how outage activity changes across the study period, while the monthly heatmap highlights seasonal and intra-annual variation. The average severity trend further shows that outage frequency and outage impact do not always evolve identically, motivating the use of severity-aware persistence measures for constructing simulation-ready outage scenarios.

**Table 1. Dataset variables and role in the analysis.**

| Variable | Description | Spatial Scale | Temporal Scale | Role in Analysis |
|---|---|---|---|---|
| County | County associated with the outage record | County | — | Primary spatial unit |
| State | State associated with the county | | — | State-level normalization and robustness analysis |
| Latitude / Longitude | Geographic coordinates of outage location or county centroid | County | — | Spatial clustering and mapping |
| Outage duration | Duration of outage event | County | Event / yearly aggregate | Severity computation |
| Customers affected | Number of customers affected by outage event | County | Event / yearly aggregate | Severity computation |
| Date | Date of outage record | — | Daily / event-level | Temporal indexing |
| Month | Month extracted from date | — | Monthly | Seasonal and heatmap analysis |
| Year | Year extracted from date | — | Annual | HPI computation |
| Season | Calendar season derived from month | — | Seasonal | Seasonal outage comparison |
| Severity score | Duration × customers affected | County | Event / annual aggregate | Hotspot ranking and scenario severity |

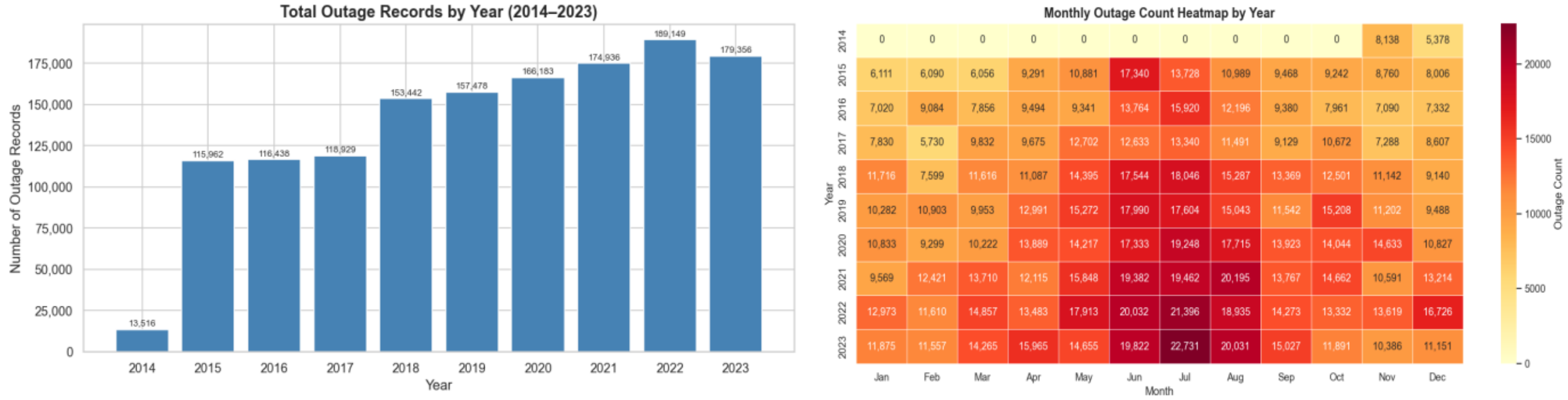

**(a) total outage records by year** **(b) monthly outage count heatmap by year**

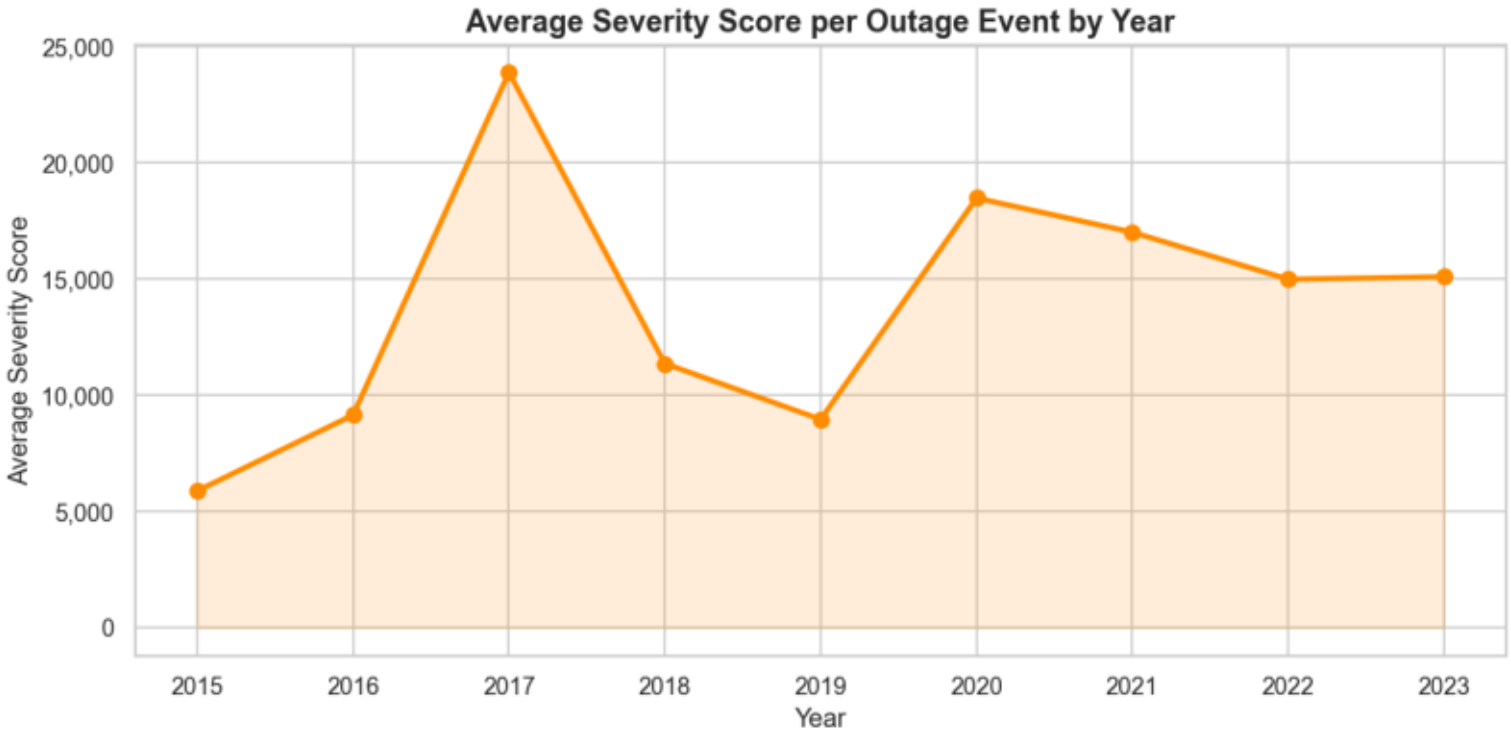

**(c) Average severity score per outage event by year**

**Figure 2: Temporal characteristics of the outage dataset.**

# 4 Hotspot Persistence Index

The hotspot persistence analysis distinguishes counties that experience isolated high-impact outage records from counties that repeatedly emerge as severe outage hotspots over multiple years. A county with one severe outage year may reflect a single extreme event, while repeated hotspot status may indicate recurring exposure, chronic infrastructure vulnerability, or repeated interaction between climate hazards and local grid conditions. To capture this distinction, we define the Hotspot Persistence Index (HPI) as a severity-ranked recurrence metric computed at the county level.

## 4.1 Annual Hotspot Definition

For each county $c$ and year $t$, outage records are first aggregated into an annual county-level severity score. As defined in Section 3, outage severity is computed as the product of outage duration and maximum customers affected. Let $S(c,t)$ in year $t$. For each year, counties are ranked according to $S(c,t)$, and a county is labeled as an annual hotspot if its severity score falls within the top $q$% of counties for that year. The annual hotspot indicator is defined as:

$$I(c,t) = \begin{cases} 1, & \text{if county } c \text{ is in the top } q\% \text{ by severity in year } t, \\ 0, & \text{otherwise} \end{cases} \quad (2)$$

In the main analysis, we set (q=10), so annual hotspots represent the top 10% of counties by severity in each year. This threshold provides a compact set of high-impact counties while preserving enough geographic coverage to construct regional simulation scenarios. For national HPI, the threshold is applied across all analyzed counties with valid annual records from 2015–2023. Defining hotspots year by year prevents one extreme outage year from being treated as persistent vulnerability.

## 4.2 HPI Definition

The Hotspot Persistence Index measures how consistently a county appears as an annual hotspot across the study period. Let $T$ be the number of years in the analysis. Since the main analysis covers 2015-2023, $T = 9$. The HPI for county $c$ is defined as:

$$HPI(c) = \frac{1}{T}\sum_{t=1}^{T} I(c,t) \quad (3)$$

The resulting value ranges from 0 to 1. An $HPI(c) = 1$ means that county $c$ appeared as a hotspot in every year of the study period. An HPI close to 0 indicates that the county rarely or never appeared as a hotspot. Intermediate values indicate partial recurrence; for example, $HPI(c) = 0.5$ means the county appeared as a hotspot in roughly half of the analysis years.

HPI is different from outage frequency or total outage severity. Frequency measures how often outage records occur, and severity measures outage magnitude. HPI measures the recurrence of high-severity status over time. Because HPI captures persistence rather than severity magnitude itself, cluster-level mean severity is retained separately and used later to parameterize simulation scenarios.

Figure 3 summarizes the county-level HPI results. Panel (a) shows the distribution of HPI values across counties, while Panel (b) maps county-level HPI values across the United States. Counties marked with stars represent locations with ($HPI = 1.0$), meaning they appear as hotspots in every year from 2015 to 2023.

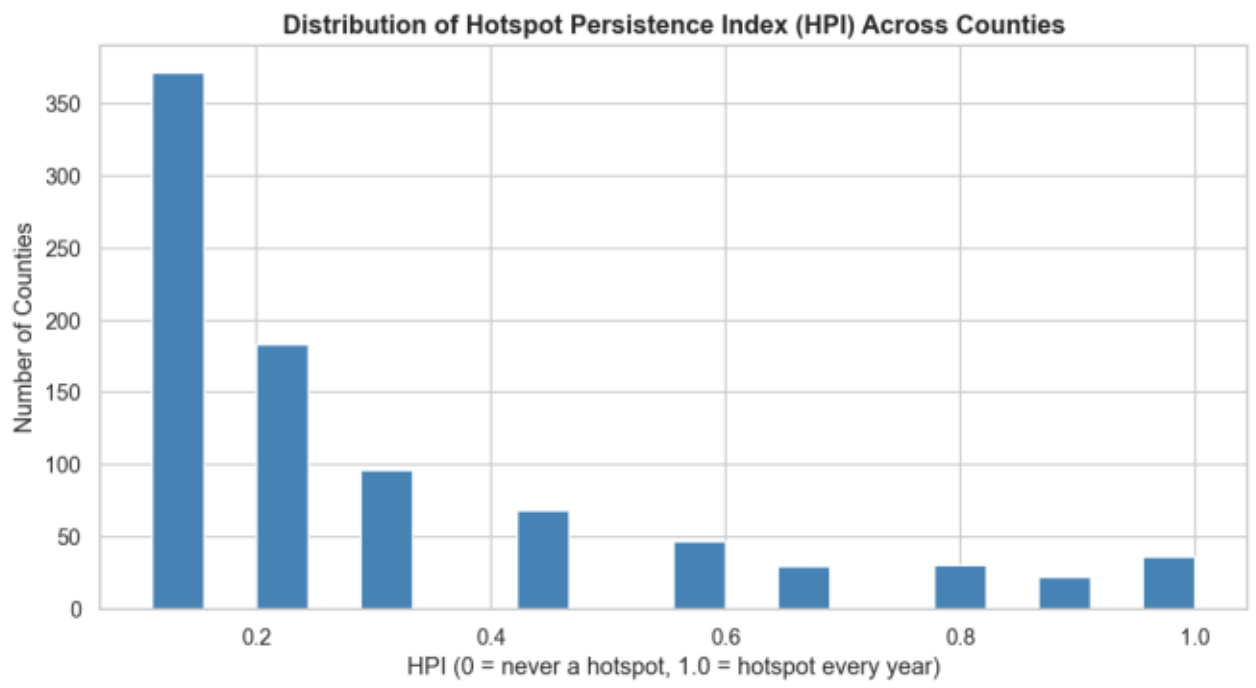

**(a): Distribution of HPI values across counties**

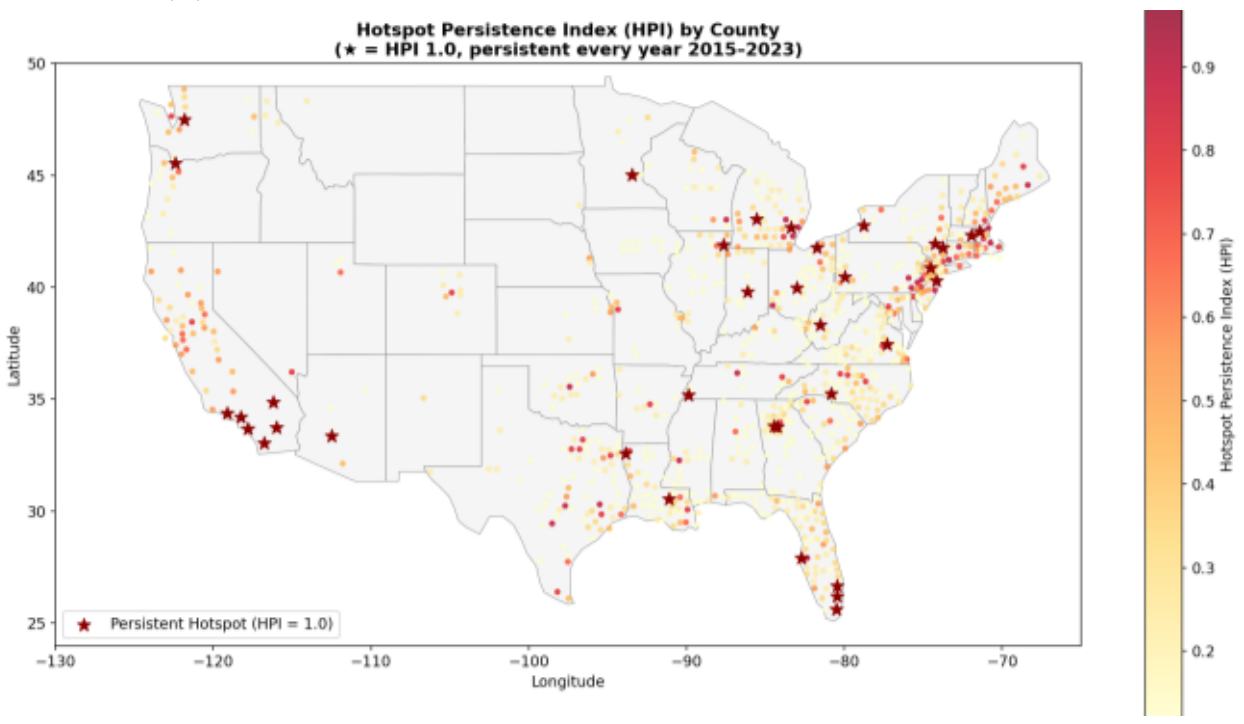

**(b): Spatial distribution of county-level HPI values**

**Figure 3: County-level Hotspot Persistence Index.**

## 4.3 National and State-Level HPI

A national hotspot definition captures counties with the largest absolute outage burden, but it may underrepresent recurrent vulnerability in smaller or less densely populated states. To address this, we compute HPI using two complementary rankings: national HPI ranks counties against all analyzed U.S. counties in each year, while state-level HPI ranks counties only against other counties within the same state using the same top-10% threshold. The overlap between the two definitions provides a robustness check. Counties that are persistent under both national and state-level definitions are treated as high-confidence hotspots because they are severe both nationally and within their local state context. In our analysis, this comparison identifies 16 doubly persistent counties, which are carried forward as robust markers of recurring outage vulnerability in the clustering and MIIM-based cascade simulation stages. Thus, HPI links county-level outage persistence to downstream regional scenario construction and simulation-based resilience evaluation.

## 5 Multi-Scale DBSCAN Scenario Construction

After computing county-level HPI values, we group outage-prone counties into spatially coherent regional disruption scenarios for downstream cascade simulation. Individual counties are too granular to serve as simulation inputs, while a single national-scale cluster would obscure regional differences in persistence, severity, and spatial extent. The goal of this stage is therefore not simply to cluster points, but to convert persistent outage counties into geographically interpretable, simulation-ready disruption zones.

A single clustering scale is insufficient for national outage data: small ($\epsilon$) values produce localized clusters, while large ($\epsilon$) values can merge adjacent regions into chain-connected components, especially along dense corridors such as the Eastern Seaboard. We therefore use a multi-scale density-based clustering process to balance regional continuity with geographic interpretability.

### 5.1 First-Pass DBSCAN Clustering

The first pass uses county centroids as spatial inputs, with each county represented by latitude and longitude. We set (min_samples=3) so that a dense neighborhood must contain at least three nearby hotspot counties, preventing isolated counties or simple county pairs from becoming independent scenarios. Although the k-distance analysis indicates a natural elbow near ($\epsilon = 0.75$) degrees, the first pass is intended to preserve broad regional continuity; therefore, we use ($\epsilon = 1.5$) degrees and apply finer refinement in later passes. This identifies broad outage-prone regions and reveals a large connected component in the eastern United States, which is refined further rather than treated as one simulation scenario.

### 5.2 Recursive Refinement of Large Components

The largest first-pass component connects several outage-prone areas in the eastern United States, including the Great Lakes, Gulf Coast/Deep South, Southeast Piedmont, Appalachian corridors, and Eastern Seaboard. We apply a second pass with ($\epsilon = 0.75$) degrees to separate this broad component into interpretable subregions. A third pass with ($\epsilon = 0.5$) degrees tests whether the remaining Eastern Seaboard structure can be further divided. It remains connected, reflecting the spatial continuity of the New England–Mid-Atlantic–Carolinas corridor, and is retained as one regional simulation scenario.

**Table 2. Final regional outage hotspot scenarios**

| Cluster ID | Region Name | County Count | Mean HPI | Mean Severity | Doubly Persistent Counties |
|---|---|---|---|---|---|
| 1 | Eastern Seaboard | 240 | 0.343 | 14314.07 | 2 |
| 2 | Gulf Coast and Deep South | 176 | 0.278 | 18770.9 | 0 |
| 3 | Great Lakes and Upper Midwest | 83 | 0.325 | 14208.94 | 4 |
| 4 | West Coast | 66 | 0.463 | 19734.28 | 2 |
| 5 | Ohio, Pennsylvania and West Virginia Corridor | 39 | 0.276 | 10986.33 | 3 |
| 6 | Southeast Piedmont | 35 | 0.292 | 11550.15 | 2 |
| 7 | Mid-South and Ozark Region | 29 | 0.249 | 14871.37 | 1 |
| 8 | Kentucky, Ohio and West Virginia Border | 17 | 0.209 | 10980.83 | 1 |
| 9 | Kentucky and Tennessee Corridor | 15 | 0.222 | 13115.86 | 0 |
| 10 | Alabama | 13 | 0.179 | 8859.96 | 0 |
| 11 | Iowa Region | 13 | 0.111 | 23302.27 | 0 |
| 12 | Central Texas | 13 | 0.299 | 16761.01 | 0 |
| 13 | Wisconsin | 9 | 0.197 | 13788 | 0 |
| 14 | Colorado and Front Range | 8 | 0.305 | 6997.09 | 0 |
| 15 | Kansas City Region | 7 | 0.365 | 12673.73 | 0 |
| 16 | Illinois and Missouri | 6 | 0.278 | 8751.41 | 0 |
| 17 | Idaho and Washington Panhandle | 6 | 0.222 | 16614.87 | 0 |
| 18 | Upper Midwest | 4 | 0.361 | 7832.64 | 1 |
| 19 | Northern Minnesota | 4 | 0.194 | 8436.6 | 0 |
| 20 | Arizona and Desert Southwest | 3 | 0.518 | 8348.99 | 0 |
| 21 | Utah and Wasatch Front | 3 | 0.296 | 12400.95 | 0 |

### 5.3 Small-Cluster Consolidation

After recursive refinement, a few very small clusters remain. We merge a small cluster only when its centroid is geographically adjacent to a larger coherent region and shares the same broader regional identity; otherwise, it is retained as an independent scenario. Noise points are not forced into clusters unless there is a clear geographic reason. This keeps the final outputs interpretable as regional outage scenarios rather than raw clustering labels.

### 5.4 Final Regional Failure Scenarios

The final multi-scale refinement process produces 21 geographically interpretable regional simulation scenarios. Of the 872 hotspot counties, 789 are assigned to final clusters and 83 remain as isolated or noise points. The scenarios capture major outage-prone regions, including the Eastern Seaboard, Gulf Coast and Deep South, West Coast, Great Lakes and Upper Midwest, Southeast Piedmont, Ohio/Pennsylvania/West Virginia corridor, Central Texas, and several smaller inland and mountain-region clusters.

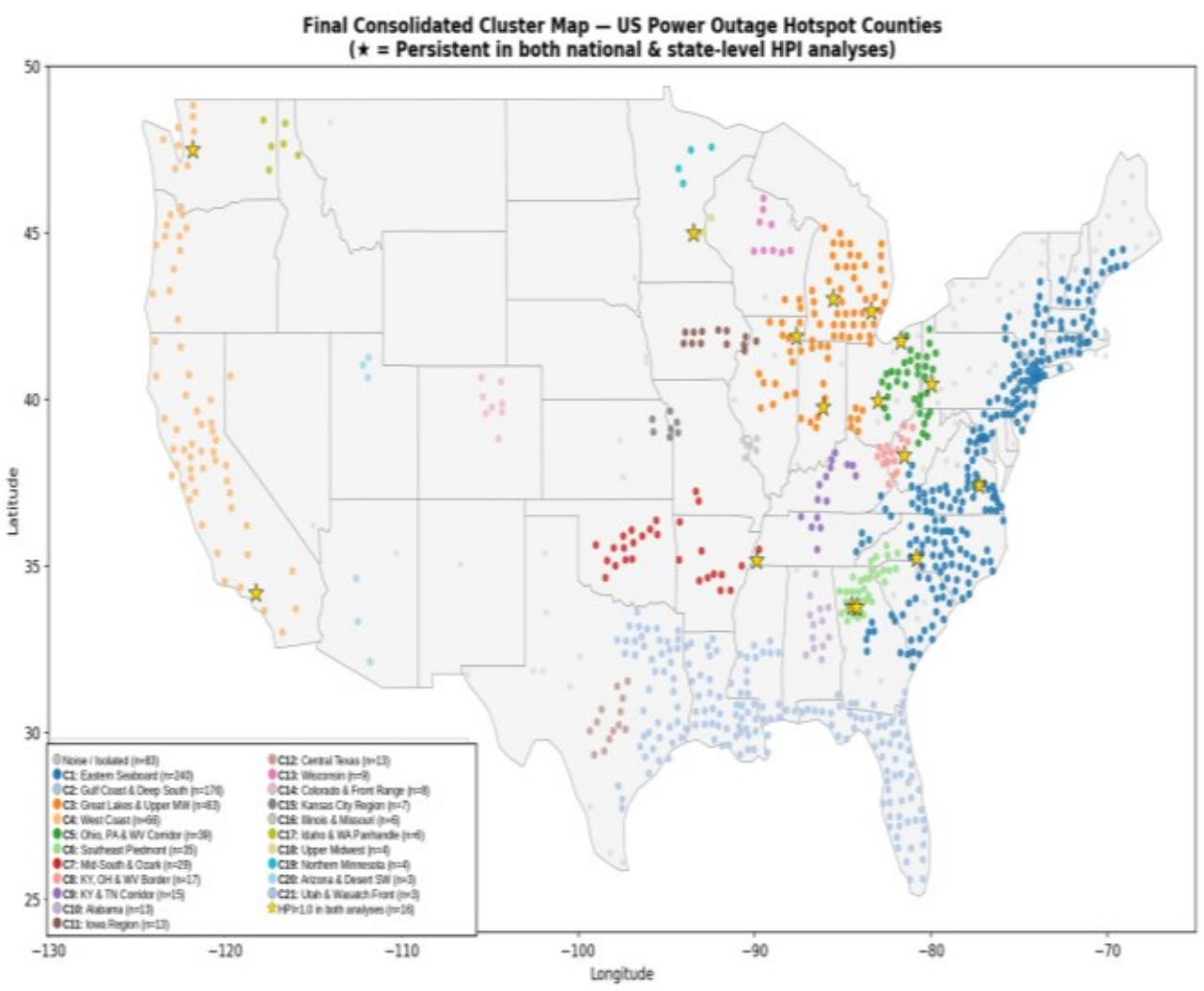

**Figure 4: Final consolidated DBSCAN cluster map of U.S. power outage hotspot counties**

Each regional scenario is summarized by region name, states represented, county count, centroid location, mean HPI, mean severity, and number of doubly persistent counties. These attributes bridge the geospatial and simulation layers: mean HPI captures persistence, mean severity parameterizes disruption magnitude, and county count defines the spatial scope of correlated initial disruptions. Figure 4 shows the final scenario map, and Table 2 summarizes the regional scenarios used in the main analysis.

## 6 Interdependent Network Cascade Evaluation

The regional scenarios constructed in Section 5 identify where outage vulnerability persists, but they do not by themselves show how disruptions propagate through interdependent infrastructure. This is especially important for power-communication systems: grid operation depends on communication infrastructure for monitoring and control, while communication entities require power to remain operational. We therefore use the geospatial scenarios as empirically derived simulation inputs and evaluate their downstream effects using the Modified Implicative Interdependency Model (MIIM) [1].

### 6.1 MIIM Overview

MIIM models a smart grid as an interdependent system composed of power entities, communication entities, and cross-layer entities. Unlike binary failure models, MIIM represents each entity using three operational states:

$$x_i = \begin{cases} 2, & \textit{fully operational state} \\ 1, & \textit{reduced operation state} \\ 0, & \textit{failed state} \end{cases} \quad (4)$$

This three-state representation is useful because empirically derived outage scenarios may correspond to either complete failure or partial degradation. MIIM captures these effects through interdependency relations (IDRs), which specify how the state of one entity depends on the states of other entities in the power and communication layers. For example, a communication entity may remain operational only if it receives power from a substation and maintains connectivity to upstream communication infrastructure. During simulation, initial failures or degradations are injected into selected entities, and MIIM iteratively evaluates the IDRs until no further state changes occur. The final entity states represent the post-cascade system condition. MIIM is suitable as a downstream simulation layer because it captures power, communication, and cross-layer dependencies while distinguishing reduced operation from complete failure.

### 6.2 Scenario Injection from Geospatial Clusters

The regional scenarios from Section 5 provide the empirical basis for MIIM simulation. We do not assume a one-to-one geographic mapping between counties and IEEE benchmark buses. Instead, the mapping is scenario-based: the IEEE 118-bus system is used as a controlled interdependency testbed for comparing empirically parameterized regional scenarios, not as a geographic representation of the U.S. grid. This design allows the same simulation substrate to be tested under different geospatially derived outage conditions.

For each cluster scenario (s), the number of initially disrupted buses is determined by the normalized county count:

$$k_s = max\{1, round((n_s / n_{max}) \times 10)\} \quad (5)$$

Here, $n_s$ is the number of counties in scenario (s), and $n_{max} = 240$ is the county count of the largest scenario, corresponding to the Eastern Seaboard cluster. The constant (10) represents the maximum number of buses injected in the largest-scope scenario. The lower bound of (1) ensures that every regional scenario injects at least one initially disrupted bus. The initial MIIM state is assigned using the normalized cluster severity score, denoted by $\sigma_s$. The normalized severity score is computed as: $\overline{\text{Severity}_s} \big/ \max_j(\overline{\text{Severity}_j})$, where $\overline{\text{Severity}_s}$ is the mean severity of scenario $s$.

$$x_i^{(0)} = 0, if \sigma_s \geq 0.70$$
$$x_i^{(0)} = 1, if \sigma_s < 0.70 \quad (6)$$

Here, $x_i^{(0)}$ is the initial state of selected entity (i) under scenario (s). State 0 denotes complete failure, and state 1 denotes degraded operation. Under this rule, the Gulf Coast/Deep South, West Coast, Iowa Region, Central Texas, and Idaho/Washington Panhandle scenarios are injected as state 0; all other scenarios, including the Eastern Seaboard, are injected as state 1. Thus, the Eastern Seaboard represents broad correlated degradation rather than complete initial failure.

For each regional scenario, we run 30 Monte Carlo simulations. In each run, $k_s$ initially disrupted buses are sampled from the IEEE 118-bus system using probability weights proportional to normalized bus betweenness centrality. The selected buses are initialized using the severity-based state rule in Eq. (6), and MIIM propagates the resulting effects until no further state changes occur. Scenario-level outcomes are averaged across the 30 runs. Thirty runs were selected to provide repeated sampling over eligible initial buses while keeping the full evaluation across regional scenarios computationally tractable. This design links the geospatial and simulation layers: county count determines disruption scope, severity determines initial state, and HPI provides the persistence weight for cascade-impact evaluation.

## 6.3 Evaluation Metrics

We evaluate each scenario using raw MIIM outcomes and HPI-weighted resilience metrics. Raw outcomes include total affected entities, complete failures/state 0, degraded entities/state 1, power-layer failures, communication-layer failures, and cascade depth. We also compute two decision-oriented metrics: HPI-weighted cascade impact and hardening gain.

$$CI_k^{HPI} = \overline{HPI_k} \times CI_k \quad (7)$$

where $CI_k$ is the average cascade impact across Monte Carlo runs, and $\overline{HPI_k}$ is the mean HPI of counties in cluster (k). This metric prioritizes scenarios that are both operationally damaging and empirically persistent. For hardening analysis, selected entities are protected by holding their state at 2, meaning fully operational. The hardening gain for entity (e) under scenario (k) is computed as:

$$HG(e,k) = CI_k - CI_k^{(e)} \quad (8)$$

where $CI_k$ is the cascade impact without hardening and $CI_k^{(e)}$ is the cascade impact when entity (e) is protected. Larger hardening gain indicates higher protection benefit. Scenario-level gains are aggregated into a single HPI-guided priority score as:

$$Priority(e) = \sum_{k \in K} Mean\, HPI_k \times HG(e,k) \quad (9)$$

where K denotes the three highest-impact regional scenarios evaluated in the hardening analysis: Eastern Seaboard, Gulf Coast/Deep South, and Great Lakes/Upper Midwest. We compare this HPI-guided priority ranking with topology-only rankings based on degree and betweenness centrality to evaluate whether empirically derived outage scenarios lead to different hardening recommendations.

**Table 3. Geospatial-to-MIIM scenario mapping rule**

| Empirical cluster attribute | MIIM mapping role | Rule |
|---|---|---|
| Mean HPI | Scenario persistence weight | Used to weight cascade impact after MIIM simulation; higher HPI gives greater influence to more persistent scenarios. |
| Mean severity | Failure magnitude | Higher-severity scenarios are initialized as complete failures/state 0; lower-severity scenarios are initialized as degraded-operation/state 1. |
| County count | Disruption scope | Larger clusters disrupt more initial MIIM entities, representing broader correlated regional stress. |
| Cluster centroid / region label | Scenario identity | Used to define the empirical regional scenario; not used as a direct county-to-bus geographic mapping. |
| Monte Carlo run | Uncertainty handling | Eligible buses/entities are sampled repeatedly according to scenario scope and severity. |

# 7 Results

This section presents the results of the proposed framework, including temporal outage patterns, hotspot persistence, regional scenario construction, MIIM cascade behavior, and hardening priorities.

## 7.1 Temporal Outage Patterns

Figure 2 summarizes temporal outage patterns in the cleaned 2015–2023 dataset. Yearly outage counts increase over the study period, with substantially higher activity after 2018; 2014 is excluded from the main analysis because coverage is incomplete. The monthly heatmap shows consistently high outage counts in June–August and elevated post-2018 activity across broader portions of the year. Figure 2(c) shows that outage severity does not always follow the same pattern as outage frequency, motivating a persistence metric based on annual severity ranking rather than frequency alone.

## 7.2 Persistent Hotspots

Figure 3 shows that county-level HPI is strongly right-skewed: most counties appear as high-severity hotspots only occasionally, while a small subset repeatedly emerges across the 2015–2023 period. Under the national top-10% annual severity definition, 36 counties achieve (HPI=1.0), meaning they rank among the most severe outage counties in every study year. State-level HPI, computed by ranking counties within each state, identifies regionally persistent vulnerability that may be underrepresented in national rankings. The overlap between national and state-level HPI identifies 16 doubly persistent counties, which are treated as high-confidence markers of recurring outage vulnerability. These results show that persistent outage risk is geographically concentrated and useful for constructing simulation scenarios.

## 7.3 Regional Cluster Structure

The multi-scale density-based clustering process produces 21 geographically interpretable regional simulation scenarios, shown in Figure 4 and summarized in Table 2. Of the 872 hotspot counties, 789 are assigned to final clusters and 83 remain as isolated or noise points. The largest scenario is the Eastern Seaboard, with 240 counties, followed by the Gulf Coast and Deep South scenario, with 176 counties and high mean severity. The West Coast scenario has one of the highest mean HPI values, while the Great Lakes and Upper Midwest scenario contains the largest number of doubly persistent counties among the major scenarios. Together, Figure 4 and Table 2 show that the final scenarios have measurable persistence, severity, and spatial scope, which are then used to parameterize MIIM cascade simulation.

## 7.4 Cascade Impact Across Regional Scenarios

Figure 5 presents MIIM cascade impact by regional scenario, ordered by HPI-weighted cascade impact and averaged across 30 Monte Carlo runs. Bars distinguish complete power-bus failures, communication-entity failures, and degraded entities. Cascade consequences are highly uneven: the three highest-impact regional scenarios account for 54.4% of total HPI-weighted cascade impact, showing that persistent geospatial vulnerability does not translate uniformly into system-level consequences.

The Eastern Seaboard and Gulf Coast/Deep South scenarios produce the largest cascade effects. The Eastern Seaboard scenario is dominated by degraded-operation outcomes, reflecting broad state-1 impact, while the Gulf Coast/Deep South scenario produces more complete failures, especially in the communication layer. The Great Lakes and Upper Midwest scenario also contributes substantially to the high-impact set used in the hardening analysis.

The Arizona/Desert Southwest scenario provides an important counterexample. It has the highest mean HPI, indicating strong recurrence of severe outage conditions, but limited MIIM cascade impact because its small spatial scope maps to a narrower correlated disruption. This shows that persistence alone is insufficient for system-level risk assessment; HPI, severity, spatial extent, and interdependency structure must be evaluated jointly.

Communication-layer entities are more vulnerable under high-impact persistent scenarios. Across these simulations, communication-layer entities fail at approximately 2.5 times the rate of power buses. This asymmetry is visible in scenarios such as Gulf Coast/Deep South and West Coast, where communication failures account for a larger share of complete failures, supporting the use of an interdependent cascade model rather than a power-only analysis.

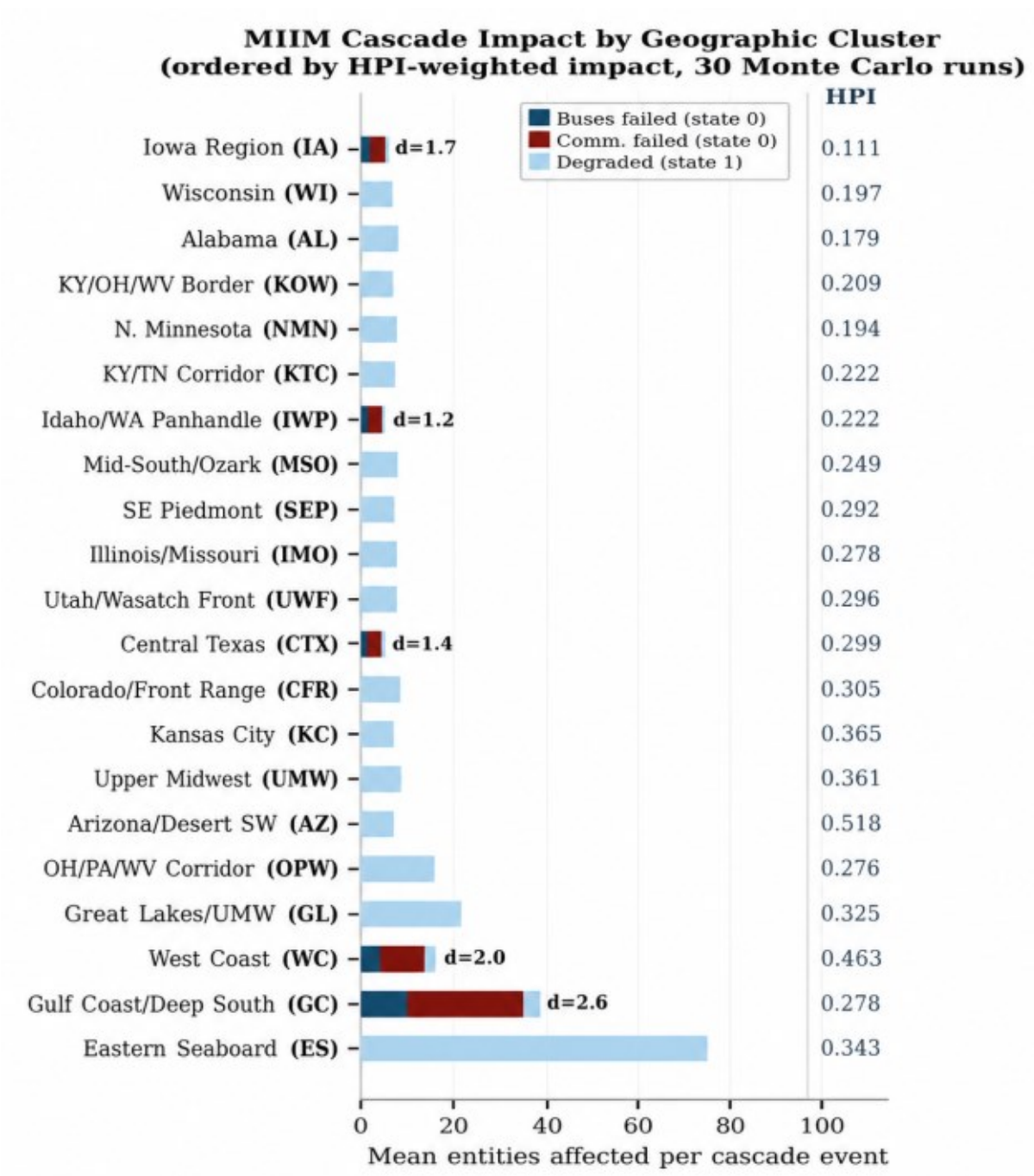

**Figure 5: MIIM cascade impact by geographic cluster**

## 7.5 Hardening Prioritization

Table 4 compares the HPI-guided hardening priority ranking from Eq. (9) with a topology-only ranking based on degree and betweenness centrality. The highest HPI-guided candidates are primarily power buses, including P49, P80, P81, P77, P68, and P69, followed by selected communication-layer entities such as G61, the gateway of substation S61. Only 2 of the top 15 rank positions match exactly between the two strategies: P49 at rank 1 and P77 at rank 4. Twelve of the fifteen entities appear in both rankings, but P81, P82, and P24 are unique to the HPI-guided list, while P37, P75, and P66 appear only under topology prioritization. Thus, HPI-guided hardening does not identify a completely different set of entities; rather, it substantially reorders protection priorities and introduces several distinct candidates based on cascade reduction under empirically persistent regional outage scenarios.

Figure 6 compares hardening gain for selected power-bus and communication entities under the Eastern Seaboard, Gulf Coast/Deep South, and Great Lakes/Upper Midwest scenarios. Hardening gain is reported in percentage points, representing the absolute reduction in the percentage of MIIM entities affected after a candidate entity is protected. Power-bus hardening produces the largest reductions, with P49 showing the strongest gain, especially under the Gulf Coast/Deep South scenario. P80 and P81 also provide substantial benefits under the Eastern Seaboard and Gulf Coast scenarios. Communication entities provide smaller but meaningful gains; gateways such as G61 and G16 indicate communication bottlenecks, while near-zero gains for some communication entities suggest that redundancy or ring-like connectivity can limit the benefit of hardening individual nodes. Overall, Figure 6 shows that persistent geospatial outage scenarios can support targeted hardening decisions rather than only identifying hotspot regions. In Figure 6, ES = Eastern Seaboard, GC = Gulf Coast/Deep South, and GL = Great Lakes/Upper Midwest.

**Table 4. HPI-guided versus topology-only hardening priorities for the IEEE 118-bus system**

| Rank | HPI-guided priority | Degree/betweenness priority | Overlap? |
|---|---|---|---|
| 1 | P49 | P49 | Yes |
| 2 | P80 | P69 | No |
| 3 | P81 | P80 | No |
| 4 | P77 | P77 | Yes |
| 5 | P68 | P65 | No |
| 6 | P69 | P68 | No |
| 7 | P65 | P100 | No |
| 8 | P100 | P38 | No |
| 9 | P17 | P30 | No |
| 10 | P30 | P70 | No |
| 11 | P23 | P37 | No |
| 12 | P70 | P17 | No |
| 13 | P82 | P75 | No |
| 14 | P38 | P66 | No |
| 15 | P24 | P23 | No |

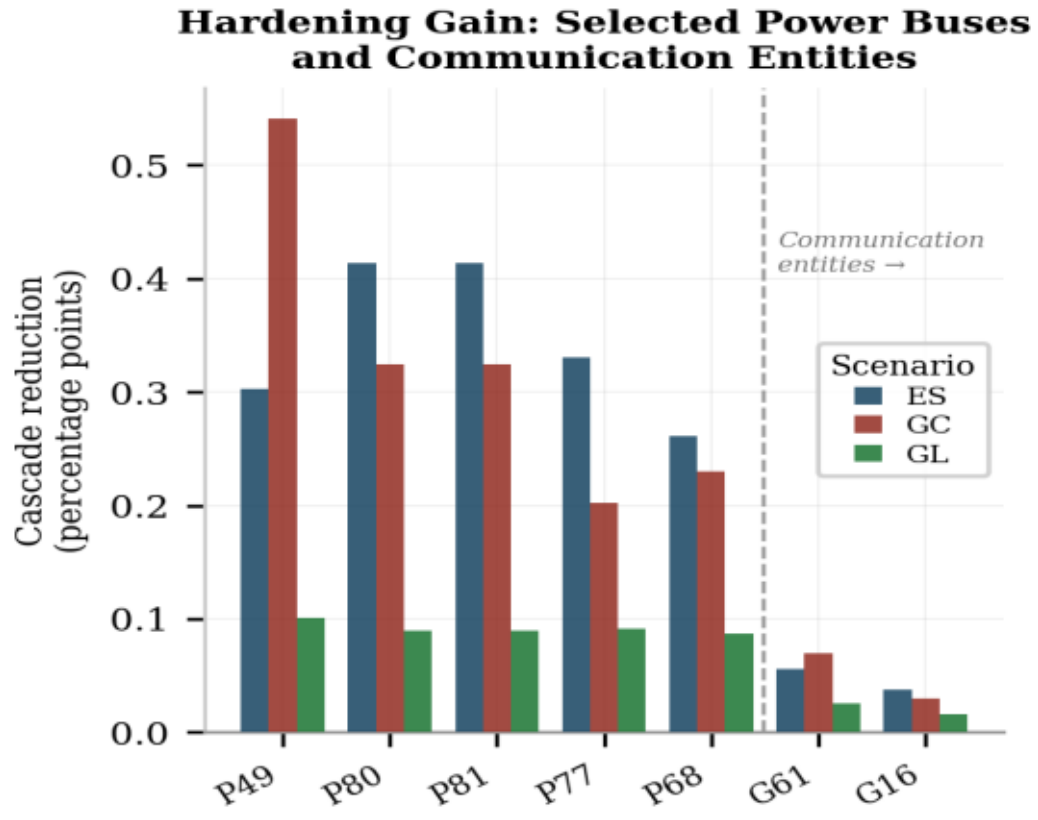

**Figure 6: Hardening gain for network entities**

## 8 Discussion and Conclusion

This paper presented a data-driven geospatial simulation framework for linking persistent power outage hotspots with downstream infrastructure resilience impacts. Using county-level outage records from 2015–2023, we introduced the Hotspot Persistence Index (HPI) to capture recurring high-severity outage status and used a multi-scale density-based refinement process to convert county-level hotspots into 21 geographically interpretable regional simulation scenarios. The national versus state-level HPI comparison identified 16 doubly persistent counties, providing a robustness check for recurring outage vulnerability.

The results show that persistent outage hotspots are not equally important from a system-resilience perspective. Some regions repeatedly appear as hotspots but produce limited cascade effects, while others with moderate HPI produce larger downstream impacts because they cover broader areas or activate more critical interdependency pathways. Three persistent regional scenarios account for 54.4% of total HPI-weighted cascade impact.

The MIIM and hardening results show why interdependency modeling matters. Several high-impact scenarios produce substantial communication-layer failures, with communication entities failing at approximately 2.5 times the rate of power buses. By weighting cascade-impact reduction according to HPI, the framework identifies entities whose protection is most valuable under persistent outage scenarios. The near-zero hardening gain observed for some ring-topology communication nodes is consistent with redundancy in communication rings, where traffic can reroute through alternate paths; this suggests that communication-layer hardening may be more effective when directed toward gateway or control-center entities that lack equivalent redundancy.

Several limitations remain. The outage dataset is observational, so the analysis identifies persistent outage patterns but does not make causal claims about specific weather events or infrastructure mechanisms. The county-level spatial unit and IEEE benchmark system support national-scale, reproducible evaluation but do not provide a direct asset-level mapping from counties to physical grid components. Therefore, the MIIM results should be interpreted as interdependency-based resilience evaluations [21] under empirically parameterized scenarios, not predictions for a specific utility system.

Despite these limitations, the framework moves outage hotspot analysis beyond descriptive mapping. Persistent outage clusters are converted into simulation-ready regional scenarios, evaluated through interdependent cascade simulation, and used to identify hardening priorities. Future work will extend this framework using utility-specific network models, asset-level outage records, weather covariates, restoration timelines, and socioeconomic vulnerability indicators to support more localized and actionable infrastructure planning.